\documentclass[aps, prd, floatfix, nofootinbib, superscriptaddress, twocolumn]{revtex4-1}

\usepackage{latexsym}
\usepackage{amsmath}
\usepackage{amssymb}
\usepackage{amsfonts}

\usepackage[mathscr,scaled=1.15]{urwchancal}
\DeclareFontFamily{OT1}{pzc}{}
\DeclareFontShape{OT1}{pzc}{m}{it}%
{<-> s * [1.15] pzcmi7t}{}
\DeclareMathAlphabet{\mathpzc}{OT1}{pzc}{m}{it}

\usepackage{color}

\usepackage{supertabular}
\usepackage{placeins}
\usepackage{epsfig}
\usepackage{graphicx}

\definecolor{purple}{rgb}{0.5,0,0.5}
\definecolor{blue}{rgb}{0.0,0,0.9}
\definecolor{prdblue}{rgb}{0.133,0.118,0.498}
\usepackage[colorlinks=true, pdfstartview=FitV, linkcolor=prdblue, citecolor= prdblue, urlcolor=prdblue]{hyperref}

\begin{document}


\title{$\,$\\[-7ex]\hspace*{\fill}{\normalsize{\sf\emph{Preprint no}. NJU-INP 023/20}}\\[1ex]
Impressions of the Continuum Bound State Problem in QCD}

\author{Si-Xue Qin}
\email[]{sqin@cqu.edu.cn}
\affiliation{Department of Physics, Chongqing University, Chongqing 401331, China.}

\author{Craig D.~Roberts}
\email[]{cdroberts@nju.edu.cn}
\affiliation{School of Physics, Nanjing University, Nanjing, Jiangsu 210093, China}
\affiliation{Institute for Nonperturbative Physics, Nanjing University, Nanjing, Jiangsu 210093, China}

\date{10 August 2020}

\begin{abstract}
Modern and anticipated facilities will deliver data that promises to reveal the innermost workings of quantum chromodynamics (QCD).  In order to fulfill that promise, phenomenology and theory must reach a new level, limiting and overcoming model-dependence, so that clean lines can be drawn that connect the data with QCD itself.  Progress in that direction, made using continuum methods for the hadron bound-state problem, is sketched herein.
\end{abstract}

\maketitle

\noindent\emph{1.$\;$History}.\,---\,%
Quantum chromodynamics (QCD) appeared a little over forty years ago as the fruit of many labours \cite{Marciano:1977su}.  The lattice formulation of the theory (lQCD) emerged contemporaneously \cite{Wilson:1974sk}; and whilst the principles and practice of perturbation theory in QCD are fairly well understood, lQCD and other nonperturbative tools are essential because the gluon and quark degrees-of-freedom used to express the Lagrangian -- QCD's partons -- are not the objects measured in detectors.

Experiments measure hadrons, each one of which is defined by its valence degrees-of-freedom.  For instance, a proton is unique in possessing two valence $u$-quarks, one valence $d$-quark, and no other valence degrees-of-freedom.  That seems simple.  However, if one chooses to work with a partonic basis, then even using a light-front formulation, with its probability interpretation, the proton wave function is so complex that no analysis has succeeded in producing anything beyond the simplest Fock space components; and information on these has only emerged recently \cite{Bali:2015ykx, Mezrag:2017znp, Bali:2019ecy}.

If the proton is such a challenge, then perhaps the pion, with its simpler structure (one valence $u$ quark and one valence $\bar d$ quark in the $\pi^+$), should be tackled first?  However, in this case, too, there are serious challenges.  For instance, the pion is a Nambu-Goldstone mode; consequently, it is unnaturally light in comparison with all other hadrons.  Hence, any realistic study of the pion must preserve chiral symmetry and the pattern by which it is dynamically broken.  These requirements have impeded many theoretical approaches using continuum methods and also challenged the lattice approach.

In the background, a route to overcoming such obstacles was identified early \cite{Pagels:1973pz, Lane:1974he, Politzer:1976tv, Pagels:1978ba, Delbourgo:1979me}.  Namely, the collection of QCD's Euler-Lagrange equations: gap equation, Bethe-Salpeter equation, Faddeev equation, \ldots, collectively known as the Dyson-Schwinger equations (DSEs) \cite{Roberts:1994dr}, can be used to solve the continuum bound-state problem.  Unsurprisingly, the earliest attempts were rudimentary; yet, they showed promise.

The DSEs are an infinite tower of coupled integral equations for a given theory's $n$-point Schwinger functions.  In QCD, the gap (Dyson) equation for the quark two-point function (propagator) involves the gluon two-point function and the three-point gluon-quark vertex.  Each of these additional Schwinger functions satisfies its own equation, which itself involves higher $n$-point functions; and so on.  To make progress in formulating a tractable problem, a truncation must be employed; and the importance of symmetries in quantum field theory means that real progress requires an intelligent scheme, \emph{i.e}.\ a systematic, symmetry-preserving method of closing the system to arrive at a finite set of equations.

\smallskip

\noindent\emph{2.$\;$Symmetry-preserving truncations}.\,---\,%
The most widely used DSE truncation method was introduced around twenty-five years ago \cite{Munczek:1994zz, Bender:1996bb}; and with a systematic approach in hand, it became possible to understand both the successes and failures of all preceding studies.

\begin{figure}[b]
\centerline{\includegraphics[clip, width=0.47\textwidth]{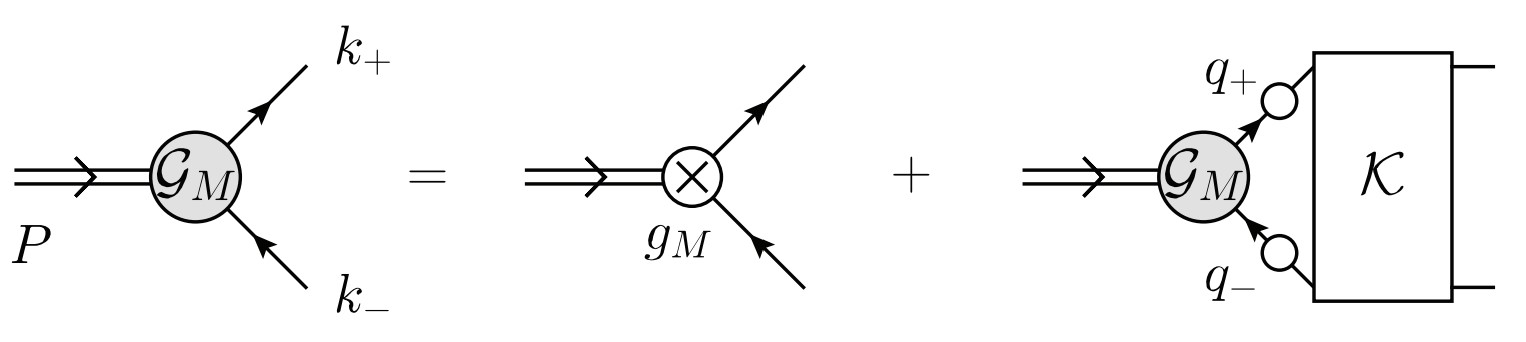}}
\caption{\label{FigIHBSE}
Inhomogeneous Bethe-Salpeter equation for a colour-singlet vertex, ${\mathpzc G}_M$, into which total momentum $P$ is flowing.  Typically, the inhomogeneity, ${\mathpzc g}_M$, is a momentum-independent matrix product that specifies the quantum numbers of the channel under consideration.
${\mathpzc K}$ is the Bethe-Salpeter kernel, which is two-particle irreducible (2PI); namely, it does not contain quark+antiquark$\,\to\,$single-gauge-boson annihilation diagrams nor diagrams that become disconnected by cutting one quark and one antiquark line.
The lines defined by open circles are dressed-quark/antiquark propagators, $S$.  Symmetries in quantum field theory entail that $S$ and ${\mathpzc K}$ are interdependent: one cannot know $S$ without knowing ${\mathpzc K}$ and vice-versa.
($k_+ = k+ \eta P$, $k_- = k - (1-\eta)P$, $\eta\in [0,1]$.  In a Poincar\'e-covariant treatment, no physical observable depends on $\eta$, the value of which defines the relative momentum.)}
\end{figure}

To be specific, consider the inhomogeneous Bethe-Salpeter equation drawn in Fig.\,\ref{FigIHBSE}, which may be written
\begin{align}
[{\mathpzc G}_M(k;P)]_{tu} & = [{\mathpzc g}_M]_{tu} \nonumber \\
& \quad +
\int_{dq}^\Lambda [\chi_M(q;P)]_{sr} {\mathpzc K(q,k;P)}^{rs}_{tu}, \label{EqBSE}
\end{align}
where $\chi_M(q;P) = S(q_+) {\mathpzc G}_M(k;P) S(q_-)$;
$r\ldots u$ denote colour, flavour and spinor indices;
and $\int_{dq}^\Lambda$ represents a translationally-invariant regularisation of the four-dimensional integral, with $\Lambda$ the regularisation scale.

The leading-order term in the scheme of Refs.\,\cite{Munczek:1994zz, Bender:1996bb} is the rainbow-ladder (RL) truncation; and with a judicious choice for the kernel, RL provides a good description of all hadronic bound-states in which (\emph{a}) orbital angular momentum does not play a significant role and (\emph{b}) the non-Abelian anomaly can be ignored.  In such circumstances, corrections to RL truncation interfere destructively.
Regarding Eq.\,\eqref{EqBSE}, the RL truncation is specified by writing ($l = k_+ -q_+= k_-  - q_-$)
\begin{subequations}
\label{EqRLInteraction}
\begin{align}
\label{KDinteraction}
\mathscr{K}_{tu}^{rs} & = {\mathpzc I}_{\mu\nu}(l) [i\gamma_\mu\frac{\lambda^{a}}{2} ]_{ts} [i\gamma_\nu\frac{\lambda^{a}}{2} ]_{ru}\,,\\
 {\mathpzc I}_{\mu\nu}(l)  & = \tilde{\mathpzc I}(l^2) T_{\mu\nu}(l)\,,
\end{align}
\end{subequations}
where $\{\lambda^a|a=1,\ldots,8\}$ are the generators of SU$(3)$-colour in the fundamental representation and $l^2 T_{\mu\nu}(l) = l^2 \delta_{\mu\nu} - l_\mu l_\nu$.  This tensor structure specifies Landau gauge, which is used because (\emph{i}) it is a fixed point of the renormalisation group; (\emph{ii}) that gauge for which corrections to RL truncation are least noticeable; and (\emph{iii}) most readily implemented in lQCD.

Evidently, Eq.\,\eqref{EqBSE} involves dressed-quark propagators; and in RL truncation, they are obtained from the following gap equation:
\begin{subequations}
\label{EqGenGap}
\begin{align}
S^{-1}(k) & = Z_2 (i\gamma \cdot k + m^{\rm bm}) + \Sigma(k)\,, \\
\Sigma(k) & = \int_{dq}^\Lambda {\mathpzc I}_{\mu\nu}(k-q)
\gamma_\mu\frac{\lambda^{a}}{2}  S(q)  \gamma_\nu \frac{\lambda^{a}}{2} \,, \label{EqSigma}
\end{align}
\end{subequations}
where $m^{\rm bm}$ is the current-quark bare mass, generated by Higgs boson couplings into QCD, and $Z_2(\zeta,\Lambda)$ is the quark wave function renormalisation constant, with $\zeta$ the renormalisation scale.

It is now straightforward to show that Eqs.\,\eqref{EqBSE}\,--\,\eqref{EqGenGap} ensure all Ward-Green-Takahashi identities \cite{Ward:1950xp, Green:1953te, Takahashi:1957xn} connected with vital global symmetries in QCD are satisfied.  Having just discussed the pion, the axial-vector identity is of particular interest here.  It can be derived from Eq.\,\eqref{EqBSE} after choosing ${\mathpzc g}_M= Z_2 \gamma_5\gamma_\mu \tau^i/2$, where $\{\tau^i | i = 1,2,3\}$ are the generators of $SU(2)$-flavour in the fundamental representation.  So long as chiral symmetry is dynamically broken, Eqs.\,\eqref{EqBSE}\,--\,\eqref{EqGenGap} are necessary and sufficient to ensure that the pion emerges as a Nambu-Goldstone mode along with all the manifold corollaries of this characteristic, \emph{e.g}.\ Refs.\,\cite{Maris:2003vk, Chang:2011vu, Chang:2013pq, Chang:2013epa, Qin:2014vya, Mezrag:2014jka, Raya:2015gva, Horn:2016rip, Li:2016dzv, Binosi:2016rxz}.

The only thing here that depends on the form of the interaction, $\tilde{\mathpzc I}(l^2)$, is dynamical chiral symmetry breaking (DCSB).  This is one of the corollaries of emergent hadronic mass (EHM) in QCD \cite{Roberts:2020udq}; hence $\tilde{\mathpzc I}(l^2)$ must be able to ensure the outcome.  More than twenty years of study have produced the following form, appropriate to RL truncation \cite{Qin:2011dd, Qin:2011xq} ($s=l^2$):
\begin{align}
\label{defcalG}
 \tfrac{1}{Z_2^2}\tilde{\mathpzc I}(s) & =
 \frac{8\pi^2}{\omega^4} D e^{-s/\omega^2} + \frac{8\pi^2 \gamma_m \mathcal{F}(s)}{\ln\big[ \tau+(1+s/\Lambda_{\rm QCD}^2)^2 \big]}\,,
\end{align}
where \cite{Qin:2018dqp, Qin:2019hgk}: $\gamma_m=12/(33-2N_f)$, $N_f=5$; $\Lambda_{\rm QCD}=0.36\,$GeV; $\tau={\rm e}^2-1$; and $s{\cal F}(s) = \{1 - \exp(-s/[4 m_t^2])\}$, $m_t=0.5\,$GeV.  Eq.\,\eqref{defcalG} preserves the one-loop renormalisation group behaviour of QCD.  Moreover, $0 < \tilde{\mathpzc I}(0) < \infty$, reflecting the fact that a nonzero gluon mass-scale appears as a consequence of EHM in QCD \cite{Aguilar:2015bud, Cui:2019dwv}.

One feature of Eq.\,\eqref{defcalG} is that when used to compute properties of light-quark ground-state vector- and flavour-nonsinglet pseudoscalar-mesons, the results are virtually insensitive to variations of $\omega \in [0.4,0.6]\,$GeV, so long as $\varsigma^3 := D\omega = {\rm constant}$.  The value of $\varsigma$ is typically chosen to reproduce the empirical value of the pion's leptonic decay constant, $f_\pi$; and in RL truncation this is achieved with $\varsigma  =0.80\,$GeV.  The quality of the description provided is illustrated by Table~\ref{obsuds}.  Comparing predictions with experiment, the absolute value of the relative error is 5(5)\%, which is typical of RL truncation within its domain of direct validity.

\begin{table}[t]
\caption{\label{obsuds}
Computed values for a range of light-quark-hadron properties (masses and leptonic decay constants), obtained using RL truncation with the interaction specified by Eq.\,\eqref{defcalG}.  The $u=d$ and $s$ current-quark masses in Eq.\,\eqref{EqGenGap} were chosen to reproduce the empirical values of $m_\pi=0.14\,$GeV, $m_K=0.50\,$GeV.  They correspond to one-loop evolved masses $m_{u,d}^{2\,{\rm GeV}}=4.8\,{\rm MeV}$, $m_{s}^{2\,{\rm GeV}}=110\,{\rm MeV}$.
(All quantities listed in GeV.)}
\begin{center}
\begin{tabular*}
{\hsize}
{|l@{\extracolsep{0ptplus1fil}}
|l@{\extracolsep{0ptplus1fil}}
l@{\extracolsep{0ptplus1fil}}
l@{\extracolsep{0ptplus1fil}}
l@{\extracolsep{0ptplus1fil}}
l@{\extracolsep{0ptplus1fil}}
l@{\extracolsep{0ptplus1fil}}
l@{\extracolsep{0ptplus1fil}}
l|@{\extracolsep{0ptplus1fil}}}\hline
       & $f_\pi$ & $f_K$ & $m_\rho$ & $f_\rho$ & $m_{K^\ast}$ & $f_{K^\ast}$ & $m_\phi$ & $f_\phi$   \\\hline
RL\,\cite{Qin:2018dqp}\; & $0.094$ & $0.11$ & $0.75$ & $0.15$ & $0.95$ & $0.18$ & $1.09$ & $0.19$ \\
expt.\cite{Zyla:2020}$\ $ & $0.092$ & $0.11$ & $0.78$ & $0.15$ & $0.89$  & $0.16$ & $1.02$ & $0.17$ \\\hline
\end{tabular*}
\end{center}
\end{table}

It will be noted that axial-vector mesons are not listed in Table~\ref{obsuds}.  In all reference frames, such systems involve significant orbital angular momentum.  RL truncation is a poor approximation in such cases because \cite{Chang:2009zb, Chang:2010hb, Chang:2011ei}: (\emph{a}) some of the corrections which interfere destructively in pseudoscalar and vector channels here interfere constructively instead; and (\emph{b}) magnifying the importance of (\emph{a}), DCSB introduces novel contributions to Bethe-Salpeter kernels that enhance spin-orbit repulsion effects.  Consequently, more sophisticated truncations are needed to describe axial-vector, scalar and tensor mesons.  In this connection, it is insufficient to improve upon RL truncation term-by-term because DCSB is essentially nonperturbative; thus, its contributions to Bethe-Salpeter kernels are missed in such a construction.  Alternatives have been developed \cite{Chang:2011ei, Williams:2015cvx, Qin:2016fwx} and are being exploited in connection with light-quark mesons.  Similarly, the $\eta$-$\eta^\prime$ complex requires an essentially nonperturbative improvement of RL truncation \cite{Ding:2018xwy}.

It is worth looking at Eq.\,\eqref{defcalG} in more detail.  Viewed as a representation of gluon exchange, the behaviour on $k^2\gtrsim (5 \Lambda_{\rm QCD})^2$ matches QCD's one-loop perturbative result; but on $k^2\in [0,(5 \Lambda_{\rm QCD})^2]$, the strength far exceeds (by up to a factor of five) the results obtained in analyses of QCD's gauge sector.  This is the cost of good phenomenological outcomes in RL truncation.  The flaw owes to the use of two bare gluon-quark vertices in Eq.\,\eqref{EqSigma}.  In reality, one of these vertices should be dressed \cite{Binosi:2016wcx}, feeding DCSB-induced enhancement into the gap equation's kernel; and that modification entails a more complicated form for ${\mathpzc K}$ in Eq.\,\eqref{KDinteraction}.  With more strength in the vertices, less is needed in the interaction.  Following development of the DCSB-improved (DB) kernel \cite{Chang:2011ei}, this is now quantitatively understood.  Namely, using the DB kernel, the gap between phenomenology and theory is bridged: the renormalisation-group-invariant effective-charge predicted by analyses of QCD's gauge sector \cite{Cui:2019dwv} is seen to be precisely the interaction required to describe hadron observables \cite{Binosi:2014aea}.  This being established, the judicious use of Eq.\,\eqref{defcalG} in RL-truncation can usefully continue because practitioners are now aware of the precise connection between their studies and QCD: the interaction's infrared overmagnification is a simple means of expressing some beyond-RL truncation effects in the predicted values of observable quantities.

\smallskip

\noindent\emph{3.$\;$Meson form factors: beyond static properties}.\,---\,%
Many qualitatively different interactions produce similar results for the spectrum of hadrons because masses are infrared-dominated quantities.  A real test of the veracity of any approach to solving QCD is its predictions for structural properties, \emph{e.g}.\ elastic and transition form factors and parton distributions.

Following the approach introduced in Ref.\,\cite{Roberts:1994hh}, the calculation of form factors is straightforward using RL truncation.  The first \emph{ab initio} prediction of the pion electromagnetic form factor, $F_\pi(Q^2)$, was described in Ref.\,\cite{Maris:2000sk}.  It was a success.  Soon after, new era experiments  \cite{Volmer:2000ek}, enabled by the continuous electron beam accelerator at Jefferson Laboratory (JLab), released results that confirmed the DSE prediction.  However, that prediction extended only to momentum transfers $Q^2 \approx 4\,$GeV$^2$ because the dressed-quark propagator obtained as a solution of the RL gap equation possesses complex conjugate poles \cite{Maris:1997tm, Windisch:2016iud} and the algorithms used in Ref.\,\cite{Maris:2000sk} were unable to handle the associated singularities moving through the integration domain sampled in the form factor calculation.

The problem was left unresolved for more than ten years; until Ref.\,\cite{Chang:2013nia} exploited the perturbation theory integral representations (PTIRs) introduced in Ref.\,\cite{Nakanishi:1969ph} and delivered predictions for $F_\pi(Q^2)$ out to arbitrarily large spacelike momenta.  These new results highlighted that QCD is not found in scaling laws; rather, it is found in scaling violations, and experiments able to reach beyond $Q^2 \approx 6\,$GeV$^2$ will see QCD scaling violations in a hard exclusive process for the first time.  This prediction has motivated new initiatives at JLab to push planned experiments to reach $Q^2 \approx 9\,$GeV$^2$ and the development of new experiments at anticipated facilities \cite{Denisov:2018unjF, Aguilar:2019teb, EicCWP, Chen:2020ijn}.  Crucially, the normalisation of $F_\pi(Q^2)$ on the domain of scaling violations is still set by EHM; hence, successful experiments will reap two rewards: validating both the modern picture of the impact of EHM on Nambu-Goldstone modes and confirmation of the predicted nature of QCD scaling violations in hard exclusive reactions.

The analysis in Ref.\,\cite{Chang:2013nia} has now been extended to a wide array of electromagnetic elastic and transition form factors of pseudoscalar mesons, including some heavy+heavy systems \cite{Raya:2015gva, Ding:2015rkn, Raya:2016yuj, Gao:2017mmp, Chen:2018rwz, Ding:2018xwy}.

RL-truncation is also challenged by mesons with a large imbalance between the current-masses of the valence quarks that define the system.  For instance, whilst the kaon does not present additional problems, the large mass difference between the $c$ and $\bar d$ quarks in the $D$ meson prevents even a direct calculation of the $D$-meson mass because of the distortion it introduces into the momentum domain that contributes in solving the Bethe-Salpeter equation.  This problem has been known for roughly thirty years; recently, a path around the difficulty was introduced.

\begin{figure}[t]
\centerline{%
\includegraphics[clip, width=0.44\textwidth]{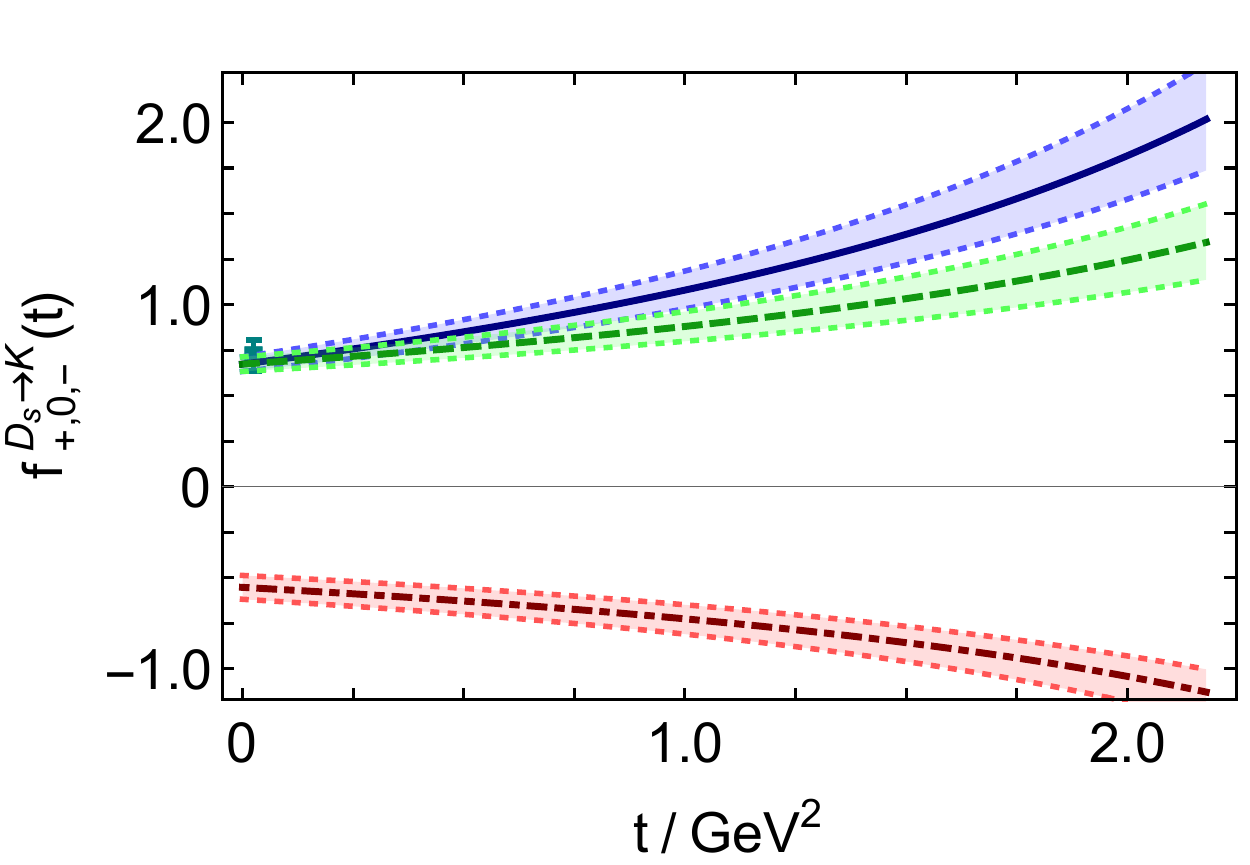}}
\caption{\label{FigDsK}
Continuum predictions for the $D_s \to K$ semileptonic transition form factors \cite{Yao:2020vef}.
Legend: $f_+$ -- solid blue curve; $f_0$ -- dashed green curve; and $f_-$ -- dot-dashed red curve.  The shaded band around each of these curves indicates the $1\sigma$ confidence level for the associated sSPM analysis.
%
Empirical datum -- cyan square \cite{Ablikim:2018upe}.
The predictions drawn here yield a branching fraction $10^3 {\mathpzc B}_{D_s^+ \to K^0 e^+ \nu_e} =
3.31(33)$ to be compared with extant experiment \cite{Ablikim:2018upe}: $3.25(38)$.  The prediction can alternatively be used to determine the CKM matrix element:  $|V_{cd}|=0.216(17)$, a value consistent with the world average, \emph{viz}.\ $0.218(4)$ \cite{Zyla:2020}.
}
\end{figure}

Naturally, one could use PTIRs.  However, that would be very time consuming if one wished to proceed beyond spectra and, \emph{e.g}.\ deliver predictions for an array of electroweak transitions.  An alternative is to employ a novel statistical implementation of the Schlessinger point method (sSPM) for the interpolation and extrapolation of analytic functions \cite{Schlessinger:1966zz, PhysRev.167.1411, Tripolt:2016cya, Chen:2018nsg, Binosi:2018rht, Binosi:2019ecz}.  Here, one exploits RL truncation to its fullest valid extent, in connection with current-quark masses and/or momentum transfers, and then uses the properties of analytic functions to extend the calculations outwards to reach the physical domains of these parameters.  This approach has been used effectively in the calculation of heavy+light meson masses, leptonic decay constants and parton distribution amplitudes \cite{Binosi:2018rht}; vector meson elastic electromagnetic form factors \cite{Xu:2019ilh}; and the leptonic and semileptonic decays of $D_{(s)}$-mesons \cite{Yao:2020vef}.

Given the prominence of related experiments at the Beijing electron+positron collider \cite{Ablikim:2015ixa, Ablikim:2018upe}, the precise results on semileptonic decays of $D_{(s)}$ mesons reported in Ref.\,\cite{Yao:2020vef} are significant.  For instance, in connection with the $D_s\to K$ transition, experiment has thus far only delivered a result for the dominant transition form factor at the maximum recoil point and there are no results from lQCD.  This is illustrated in Fig.\,\ref{FigDsK}.

There is a perspective from which one may view hadron parton distributions as even more fundamental than form factors.  They are harder to measure; but interpreted carefully, parton distributions provide local maps of the momentum and spatial distributions of partons within hadrons.  It is also far more difficult to deliver sound predictions for parton distributions. These things notwithstanding, following the first study \cite{Hecht:2000xa}, significant progress has been made using DSE methods, which may be followed from Refs.\,\cite{Ding:2019qlr, Ding:2019lwe, Cui:2020dlm}.

\smallskip

\begin{figure}[!t]
\includegraphics[clip,width=0.95\linewidth]{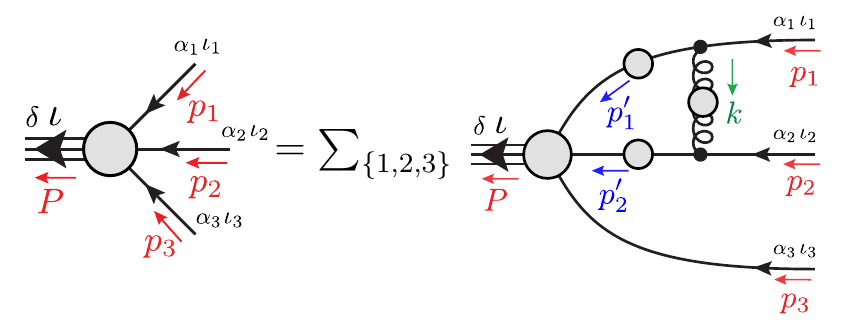}
%
\caption{\label{FEimage}
Homogeneous three-body equation used widely to compute baryon masses and bound-state amplitudes: total momentum $P=p_1+p_2+p_3$.  The equation has solutions at $P^2=-m_B^2$, where $m_B$ are the baryons masses.
Amplitude: vertex on the left-hand-side;
spring with shaded circle: quark-quark interaction kernel in Eqs.\,\eqref{EqRLInteraction};
and solid line with shaded circle: dressed-propagators for scattering quarks, obtained by solving the gap equation, Eq.\,\eqref{EqGenGap}.
}
\end{figure}

\noindent\emph{4.$\;$Baryons}.\,---\,%
The first hadron discovered was the proton, about which much is now known: not just its mass, but also a great deal of structural information.  Many more baryons have also been found.  Thus, in addition to a capacity to explain mesons, any viable continuum approach to the bound state problem in QCD must offer answers to baryon problems.  This is provided by the Poincar\'e-covariant Faddeev equation \cite{Cahill:1988dx, Burden:1988dt, Cahill:1988zi, Reinhardt:1989rw, Efimov:1990uz}.  Generalising the Bethe-Salpeter equation drawn in Fig.\,\ref{FigIHBSE} to treat systems seeded by three valence quarks, one can derive the homogeneous integral equation depicted in Fig.\,\ref{FEimage}.  This is a RL-truncation equation and the running sum ensures that the Faddeev kernel joins any two valence-quark participants once and only once.

The Faddeev equation in Fig.\,\ref{FEimage} has been used as the basis for computation of baryon masses and some elastic and transition form factors.  Many of these applications are described in Ref.\,\cite{Eichmann:2016yit}.  It has also been used to provide Poincar\'e-covariant calculations of: the proton's tensor charges \cite{Wang:2018kto}; and the spectrum of $J^P=1/2^+$, $3/2^+$ baryons, including those with heavy-quarks, and their first positive-parity excitations \cite{Qin:2018dqp, Qin:2019hgk}.

\begin{figure}[t]
\includegraphics[clip,width=0.95\linewidth]{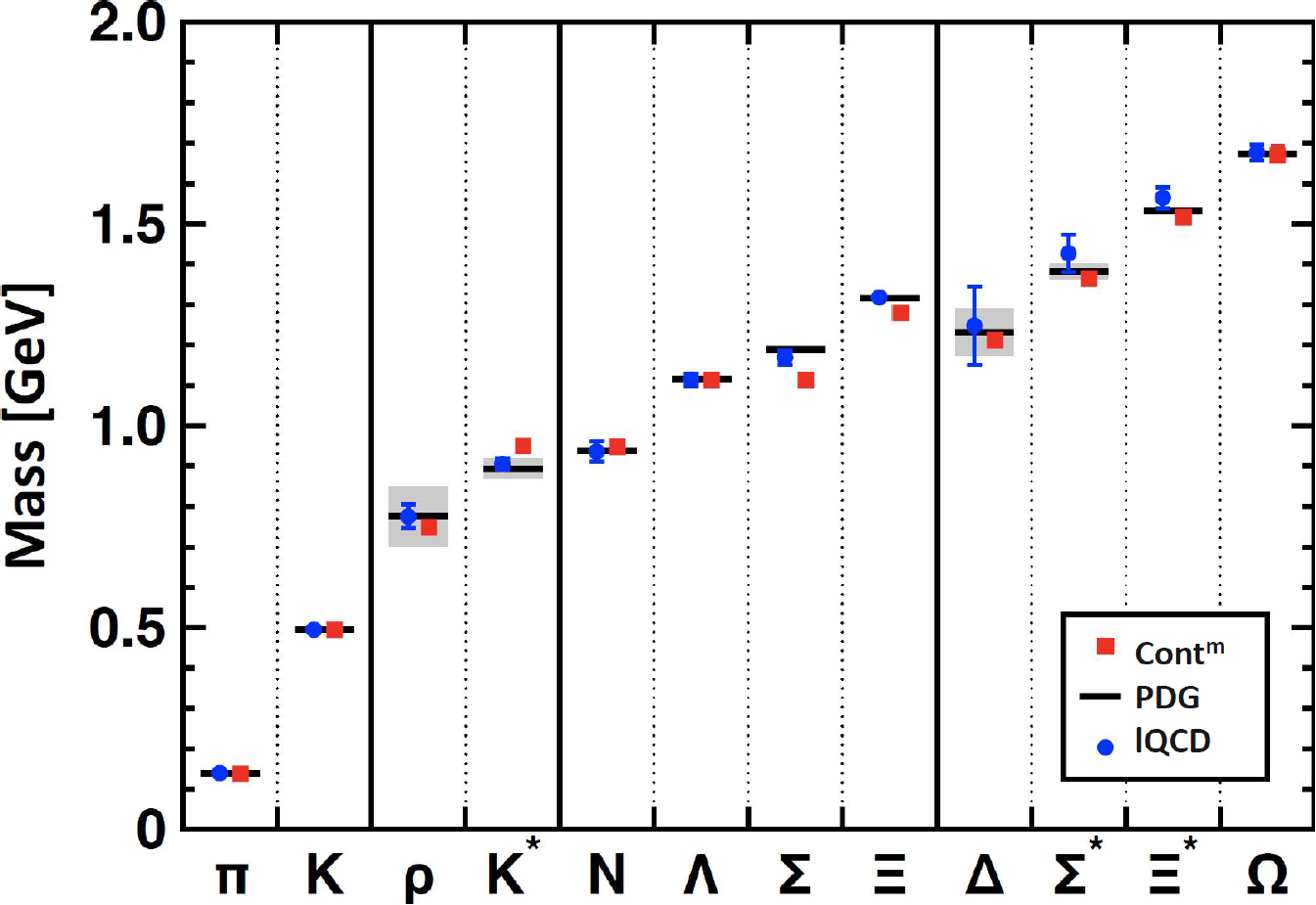}
%
\caption{\label{810MassComparison}
Masses of pseudoscalar and vector mesons, and ground-state positive-parity octet and decuplet baryons calculated using continuum (Cont$^{\rm m}$ -- squares, red) \cite{Qin:2019hgk} and lattice \cite{Durr:2008zz} methods in QCD compared with experiment \cite{Zyla:2020} (PDG: black bars, with decay-widths of unstable states shaded grey).
}
\end{figure}

The power of the approach is illustrated by Fig.\,\ref{810MassComparison}, which compares predictions made using Fig.\,\ref{FEimage}, completed using Eqs.\,\eqref{EqRLInteraction}, \eqref{EqGenGap}, with experiment and lQCD results.  The continuum predictions were made with $m_u=m_d$, in which limit RL-truncation does not distinguish between isospin $I=0,1$ systems; but such isospin symmetry breaking effects are small, as highlighted by the empirically determined $\Sigma$-$\Lambda$ mass difference ($< 7$\%).  Fig.\,\ref{810MassComparison} reveals that two disparate approaches to hadron bound states \cite{Qin:2019hgk, Durr:2008zz} produce a ground-state spectrum in good agreement with experiment.  Importantly, more states are accessible using the continuum approach \cite{Qin:2019hgk, Qin:2018dqp}.

\begin{figure}[t]
\vspace*{2.7ex}

\leftline{\hspace*{2em}{\large{\textsf{A}}}}
\vspace*{-5ex}
\centerline{\includegraphics[clip, width=0.42\textwidth]{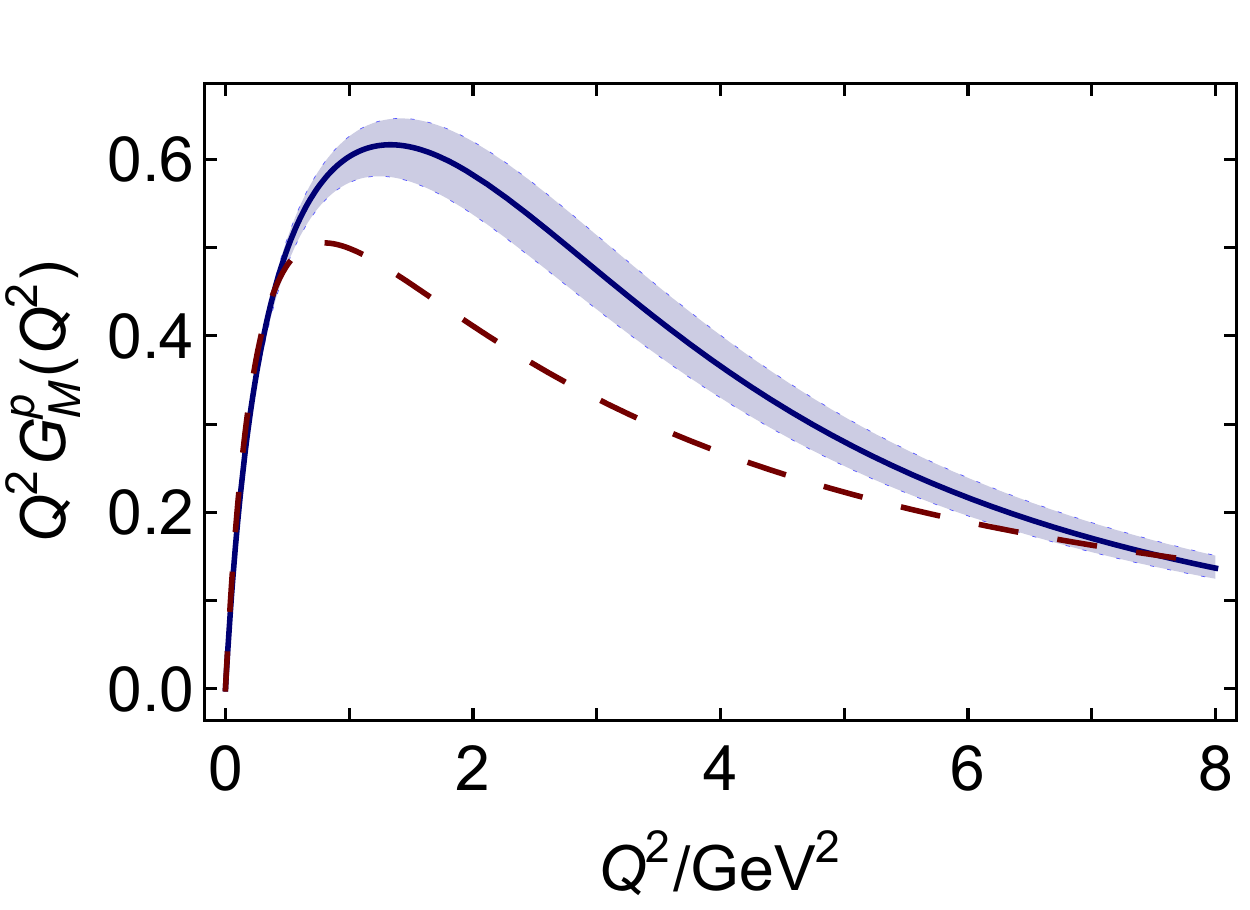}}
\vspace*{2ex}

\leftline{\hspace*{2em}{\large{\textsf{B}}}}
\vspace*{-5ex}
\centerline{\includegraphics[clip, width=0.42\textwidth]{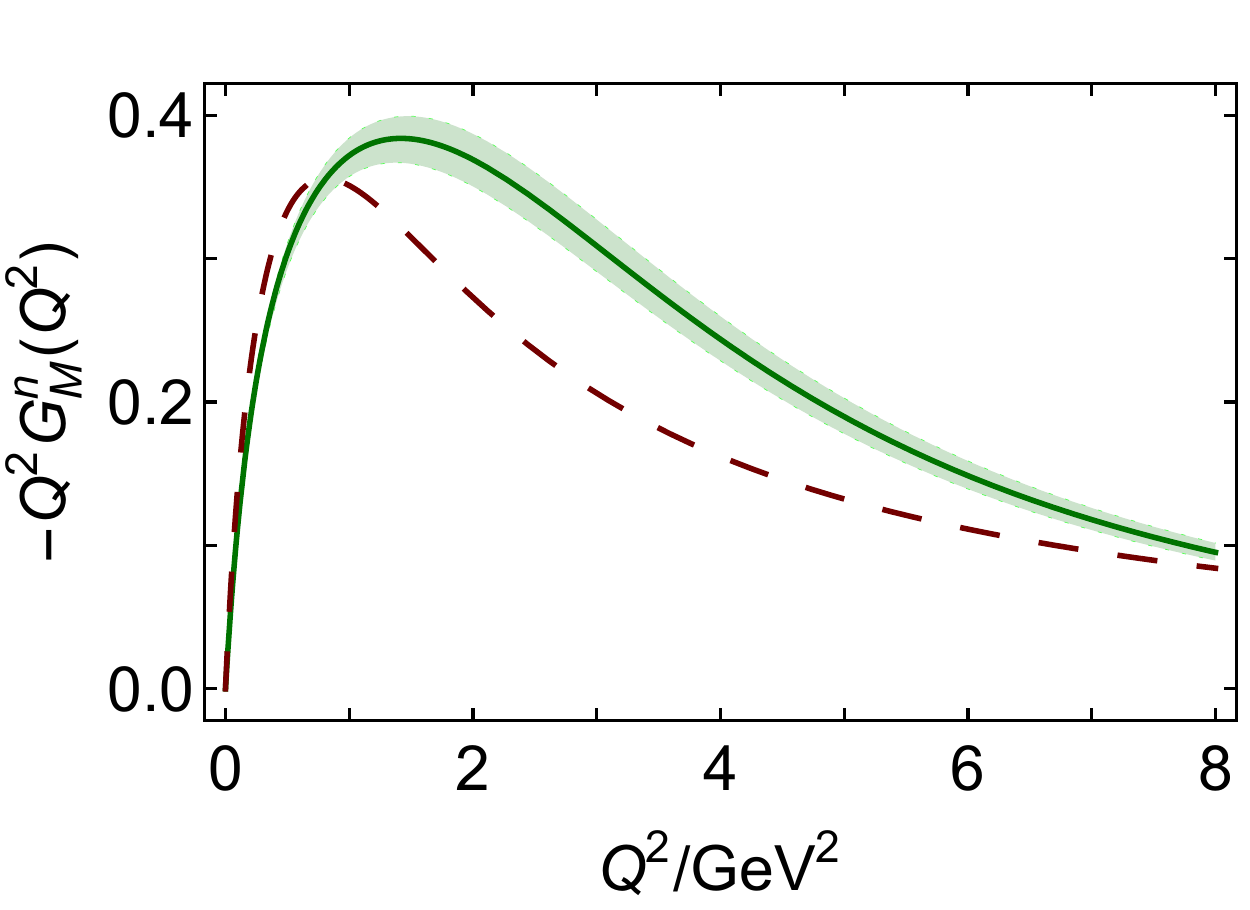}}
\caption{\label{SXQGMN}
Sachs magnetic form factors computed using the interaction in Eq.\,\eqref{defcalG}.
(\textbf{A}) Proton: solid blue curve.
(\textbf{B}) Neutron: solid green curve.
Both panels: bands bracketing the solid curves show the effect of $\omega \to 0.5 (1 \pm 0.1)\,$GeV in Eq.\,\eqref{defcalG}; and the long-dashed red curves are the parametrisations of empirical data in Ref.\,\cite{Kelly:2004hm}.
}
\end{figure}

Having obtained a baryon's bound-state amplitude by solving the equation in Fig.\,\ref{FEimage}, one can use the associated interaction current \cite{Eichmann:2011vu} to compute form factors.  As an example, in Fig.\,\ref{SXQGMN} we depict the first \emph{ab initio} results for nucleon Sachs magnetic form factors obtained using the interaction in Eq.\,\eqref{defcalG}.  Namely, the results were obtained by beginning with this interaction, using RL truncation to compute all elements entering into the equation depicted in Fig.\,\ref{FEimage}, solving that equation to obtain the nucleon Faddeev amplitudes, and subsequently combining all elements with the Poincar\'e-covariant current to calculate the form factors.  In calculating the results in Fig.\,\ref{SXQGMN}, an algorithm similar to that used for $F_\pi(Q^2)$ in Ref.\,\cite{Maris:2000sk} was employed.  Consequently, the same problem was encountered; hence, calculation was impossible beyond $Q^2 \approx 6\,$GeV$^2$.  Pad\'e approximants of order $[1,3]$ were used to extrapolate onto $Q^2/{\rm GeV}^2 \in [6,8]$.  Analyses underway will exploit the sSPM to extend these predictions onto the entire domain of momentum transfers accessible at the upgraded JLab \cite{Gilfoyle:2018xsa, Wojtsekhowski:2020tlo}.

For comparison, Fig.\,\ref{SXQGMN} also displays parametrisations of empirical data \cite{Kelly:2004hm}.  (Updated parametrisations are available, but they are not materially different.)  Evidently, there is fair agreement between the calculated results and experiment; but mismatches are noticeable on $Q^2 \lesssim 3\,$GeV$^2$.
Some part of the discrepancies can be understood as follows.  The RL truncation produces a hadron's dressed-quark core, \emph{viz}.\ those contributions to the DSE kernels which might properly represent meson-baryon final-state-interactions (the meson cloud) are suppressed \cite{Hecht:2002ej}.  Such effects are soft, so their impact is restricted to the low-$Q^2$ domain.  Hence, the mismatches owe, at least in part, to omission of the nucleon's meson cloud.
Exactly how much is not yet known and it varies between systems; so many different paths are being followed in studies that are underway to learn more, \emph{e.g}.\ Refs.\,\cite{Burkert:2017djo, Chen:2017mug, Williams:2018adr, Lu:2019bjs} and references therein.

The RL truncation itself may also be partly at fault.  Numerous studies in experiment and theory have shown that nonpointlike, fully-interacting quark+quark correlations (diquarks) are very likely important in understanding baryon structure \cite{Brodsky:2020vco, Barabanov:2020qq}.  Such correlations typically introduce marked SU$(6)$ spin-flavour symmetry breaking in baryon wave functions.  RL truncation does produce some SU$(6)$ breaking, enforcing couplings between wave function components in momentum-, spin- and flavour-spaces, and it contains the seeds for the formation of diquark correlations.  However, RL truncation does not, \emph{e.g}.\ produce the mass-splittings amongst isospin partners that are achieved in quark+diquark truncations of the Faddeev equation.  Therefore, an important contemporary challenge is to find a practicable beyond-RL truncation of the Faddeev equation that can produce diquark correlations and then exploit it to best effect.

\smallskip

\noindent\emph{5.$\;$Prospects}.\,---\,%
The bound-state problem in QCD has existed for over forty years.  During the first half of this period, some important progress was made with continuum methods, but it was unclear whether they could move much beyond computation of the static properties of the lightest mesons.  The picture changed rapidly in the following vicennium.  Now, DSE predictions are available for a wide variety of observables; they explain much existing data and support, guide and stimulate experiments at many of the world's leading accelerator facilities \cite{Denisov:2018unjF, Aguilar:2019teb, EicCWP, Brodsky:2020vco, Chen:2020ijn, Barabanov:2020qq}.  These roles are crucial as high-energy nuclear and particle physics enters a new era, with the operation, construction, and planning of high-luminosity, high energy facilities that promise to see deep into the heart of hadrons with unprecedented clarity.

We have given a few impressions of recent progress with the continuum bound-state problem in QCD.  Many others are worthy of mention, \emph{e.g}.:
studies of in-medium QCD \cite{Fischer:2018sdj, Gao:2020qsj};
contributions to unravelling the puzzle surrounding the anomalous magnetic moment of the muon \cite{Raya:2019dnh, Eichmann:2019bqf};
and
analyses of potentially hybrid/exotic systems, \emph{i.e}.\ states whose valence content is not simply $q\bar q$ or $qqq$, which can be tackled using generalisations of the bound-state equations described above \cite{Xu:2019sns, Souza:2019ylx, Lucha:2019cpe, Kaptari:2020qlt, Wallbott:2020jzh, Huber:2020ngt}.

Regarding the future, attention will focus on delivering \emph{ab initio} predictions for generalised- and transverse-momentum-dependent (TMD) parton distributions because (\emph{a}) experiments aimed at measuring these quantities have a high profile worldwide; (\emph{b}) the data obtained will be of limited worth unless there are QCD-connected predictions that can be used to understand them; and (\emph{c}) so long as (\emph{b}) is achieved, it will become possible to draw realistic three-dimensional images of hadrons.

The challenge of understanding TMDs brings with it the need to calculate and comprehend parton fragmentation functions (PFFs).  PFFs describe how partons, generated in a high-energy collision and (nearly) massless in perturbation theory, convert into a shower of massive hadrons, \emph{i.e}.\ PFFs describe how hadrons with mass emerge from massless partons.  As such, PFFs are intimately connected with the problem of confinement in empirical QCD, \emph{viz}.\ the QCD with light-quark degrees-of-freedom that is at work in our Universe.  In our view, confinement, dynamical chiral symmetry breaking (DCSB) and emergent hadronic mass (EHM) are three sides of the same triangle, with EHM at the base; and an understanding of PFFs can potentially make this manifest.

\smallskip

%
\noindent\emph{Acknowledgments}.\,---\,%
We are grateful for constructive comments from D.~Binosi, L.~Chang and Z.-F.~Cui.
Work supported by:
Jiangsu Province \emph{Hundred Talents Plan for Professionals}.
%


\end{document}